\documentclass[ukenglish]{nik}
\usepackage{mathptm}
\usepackage{cite}
\usepackage{graphicx, epstopdf}
\usepackage{pgfplots}
\usepackage{amsmath}
\usepackage[caption=false,font=footnotesize]{subfig}
\usepackage{listings}
\usepackage{pgfplotstable}
\usepackage{url}
\usepackage{authblk}

\begin{document}
\title{Clustering Methods for Electricity Consumers: An Empirical Study in Hvaler - Norway}

\author[1]{The-Hien Dang-Ha\thanks{hthdang@student.matnat.uio.no}}
\author[2]{Roland Olsson\thanks{roland.olsson@hiof.no}}
\author[3]{Hao Wang\thanks{hawa@ntnu.no}}
\affil[1]{Department of Informatics, University of Oslo}
\affil[2]{Faculty of Computer Sciences, Ostfold University College}
\affil[3]{Faculty of Engineering and Natural Sciences, NTNU}

\date{10 August 2016}

\maketitle
\begin{abstract}
	The development of Smart Grid in Norway in specific and Europe/US in general will shortly lead to the availability of massive amount of fine-grained spatio-temporal consumption data from domestic households. This enables the application of data mining techniques for traditional problems in power system. Clustering customers into appropriate groups is extremely useful for operators or retailers to address each group differently through dedicated tariffs or customer-tailored services. Currently, the task is done based on demographic data collected through questionnaire, which is error-prone. In this paper, we used three different clustering techniques (together with their variants) to automatically segment electricity consumers based on their consumption patterns. We also proposed a good way to extract consumption patterns for each consumer. The grouping results were assessed using four common internal validity indexes. We found that the combination of Self Organizing Map (SOM) and k-means algorithms produces the most insightful and useful grouping. We also discovered that grouping quality cannot be measured effectively by automatic indicators, which goes against common suggestions in literature.
\end{abstract}
\section{Introduction}
The electricity industry in Norway is witnessing a revolution where many \textit{Smart Grid} concepts are being developed and experimented in several large-scale demo sites linked to real power systems with thousands of real consumers \cite{fosso2014moving}. The revolution is strongly driven by the regulator, e.g., it is required that smart meters be installed for all electricity customers by 2019-01-01 \cite{fosso2014moving}. This will shortly lead to the availability of a massive amount of fine-grained spatio-temporal consumption data from domestic households. This data will enable the application of state-of-art technologies for traditional problems in various grid operations. In this paper, we focus on consumer classification in smart grid. Clustering consumers into appropriate groups is extremely important for distribution system operators and retailers, as they can address each group differently through dedicated tariffs, energy consulting, or other consumer-tailored energy services \cite{Chicco2006, Beckel2012, Chicco2012}. 

Currently, the utility companies usually group customers based on demographic data (family size, house size, location, number of appliances, heating technology etc.). The method is error-prone since the data is usually collected through questionnaire when the electricity connection is first built, and hardly ever updated. Moreover, demographic data usually does not correlate well to the actual evolution of the electricity consumption.

In addition to these problems, the power grid in Norway also has unique consumption patterns. There were $419,246$ holiday houses or cabins in Norway as of January 2015 \cite{ssb}, since Norwegian has a long tradition of second-home tourism \cite{Vitterso2007}. These cabins are only occupied during weekends, holidays, or warmer seasons. Hvaler is a good example. Being a small island located approximately 120km south of Oslo, Hvaler has a permanent population of only around $7000$ \cite{bremdal2014seasonal}. However, during summer or vacation, the population at Hvaler may reach more than $30,000$. Therefore, data collected from Hvaler is suitable for evaluating performance of automatic clustering methods, especially on their ability to distinguish between normal households and cabins. The dataset used in this paper contains 3090 hourly consumption time-series, collected from Hvaler during the years 2012-2013. Although all of those time-series are currently recorded in our database as ``Household'', applying appropriate automatic clustering methods reveals the fact that the dataset indeed contains cabins, street lightings, and other misclassified consumers. This proves the superior ability of automatic clustering over traditional approach. 

To evaluate and compare performance of clustering algorithms, we used four clustering validity indicators. Those indicators are commonly used in literature to select the best clustering method or choose the optimal cluster size \cite{Tsekouras2007, Wijaya2014, Figueiredo2005, Chicco2003}. However, experiment result in this paper shows that these indicators are strongly biased (at least in our dataset) and should be used with great care.

The remainder of this paper is structured as follows. Section \ref{related} presents a short review of different approaches on electricity customer classification. Section \ref{feature} shows how the representative load patterns are designed and extracted. Section \ref{sec:validity} discusses how we can measure quality of clustering results. Section \ref{clustering} explains briefly all the chosen clustering techniques and their variants. Section \ref{results} presents and discusses all experiment results, and Section \ref{conclusion} concludes the paper.

\section{Related Work}
\label{related}
Customer classification has been studied extensively with many proposed techniques and tools. In this section, we generally classified them into the following three main approaches, depending on the types of features on which the analysis is based.

\begin{itemize}
	\item \textit{Intrinsic clustering:} uses features that are extracted directly from consumption time-series. These features could be statistics of the time-series (e.g. total usage, maximum usage, standard deviation at peak hour) \cite{Dent2012}, load shape indexes (e.g. load factor, night impact, lunch impact) \cite{Figueiredo2005}, frequency domain indexes (e.g. harmonics amplitude obtained by Discrete Fourier Transform) \cite{Verdu2004}. However, the most popular approach is to use \textit{Representative Load Pattern} (RLP). For a given customer, a RLP is calculated by averaging and normalizing the load data corresponding to a specified loading condition \cite{Chicco2003, Chicco2006, Chicco2012, Tsekouras2007, Mutanen2011, Hernandez2012}. 
	\item \textit{Extrinsic clustering:} uses external features that are extracted from weather or economics variables. However, these features are not guaranteed to always well correlate with the consumption patterns. Therefore, the combination of extrinsic and intrinsic approach are usually employed. 
	
	\item \textit{Hybrid clustering:} uses both intrinsic and extrinsic features. A two-stage method is usually applied, where in the first stage, an initial clustering is conducted based on extrinsic features (e.g. is it a residential, industrial, commercial or lighting). The intrinsic features are then employed in the second stage to further classify within these macro-categories \cite{Verdu2004}. Another hybrid approach is to first remove the effect of extrinsic features (e.g. temperature) from consumption data before applying intrinsic clustering \cite{Ardakanian2014}. Or we could use extrinsic features together with consumption data to reveal occupancy states \cite{Albert2013}, thermal profiles \cite{Albert2013Thermal}, or demographic information \cite{Beckel2012}.
\end{itemize}

In this paper, we followed the RLP approach, which does not require external features. We believe that consumers are well-characterized by their consumption behaviors so that a good clustering method should be able to classify them basing solely on their consumption data.

\section{Extracting Representative Load Pattern}
\label{feature}
Feature extraction is the most important step in the whole clustering process, as it will decide usefulness of the grouping result. The chosen features must provide suitable and enough information to distinguish between groups that we need to cluster, as well as keep the noise or irrelevant information low so that the clustering algorithms could work well. Hand-designed features such as statistics or load shape indexes are less noisy, but they can be biased by the designer's assumptions about how the groupings should be. Therefore, using hand-designed features can cause some consumer classes to be indistinguishable from others.

\begin{figure}
	\centering
	\includegraphics[width=0.7\linewidth]{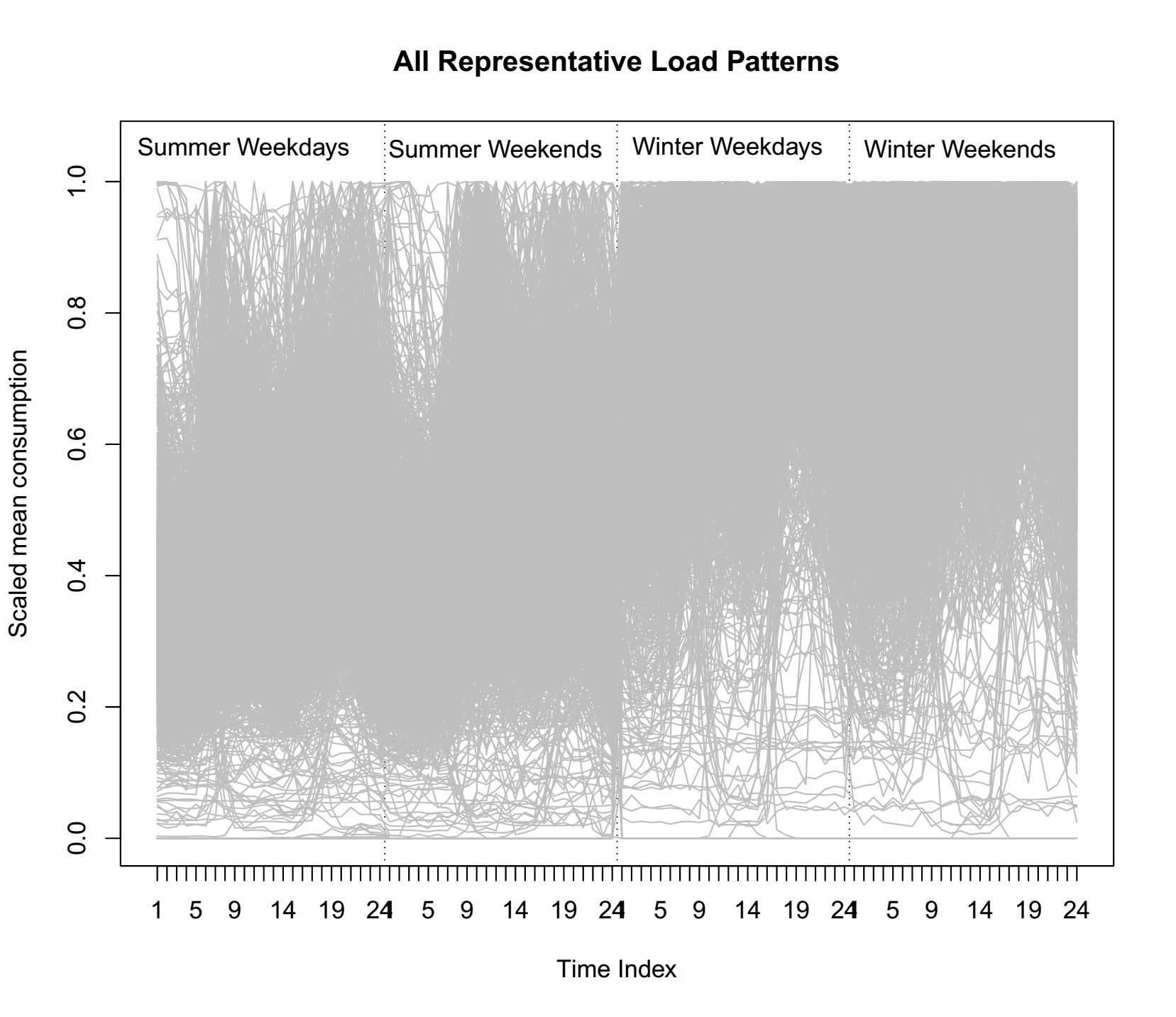}
	\caption{All 3090 Representative Load Patterns in Hvaler dataset. Each line is RLP of one consumer}
	\label{fig:AllRLP}
\end{figure}

The representative load pattern (RLP) approach handles this problem by representing each consumer by his daily consumption pattern under different contexts. In this paper, we consider four contexts: summer weekdays, summer weekends, winter weekdays, and winter weekends~\footnote{We consider all public holidays in Norway as weekends}). These contexts are designed to provide enough information to differentiate important consumer classes such as industrial consumers, normal households, or cabins. In each context, there are $24$ features, each is the average consumption during a particular hour of day for all days satisfying the context. For example, the first RLP feature of a consumer is his average consumption at $0$AM for all summer weekdays. Similarly, since we have $24$ features ($24$ hours) for each context and $4$ contexts, a RLP has $96$ features in total. To make the RLP scale-invariant, all the RLPs are normalized to the $[0, 1]$ range by dividing by its maximum value. Figure \ref{fig:AllRLP} shows all 3090 RLPs of Hvaler dataset, each line is RLP of one consumer. Ones can see that although all the RLPs are currently recorded as households in our database, their consumption patterns are very different. The job of clustering methods is to group those RLPs into compact groups with distinct and meaningful properties.

\section{Validity Indexes for Assessing Clustering}
\label{sec:validity}
Before diving into clustering methods, first we need to discuss how we measure the quality of clustering results.
\subsection{The Clusters' Usefulness}
The real usefulness of a cluster partition is hard to measure and depends on specific application. However, as suggested in \cite{Dent2012,sarstedt2014concise}, there are several common criteria that a cluster should meet. The criteria relevant to our clustering task are: (1) \textit{Compactness}: the groupings must be homogeneous within and heterogeneous between each other; (2) \textit{Differentiable}: the groupings must be distinguishable conceptually, and respond differently to different potential programs; (3) \textit{Substantial}: the groupings are large enough for a particular program benefits from; (4) \textit{Stable}: the groupings should be stable over time to be worth designing dedicated programs for; (5) \textit{Actionable}:  actions / programs can be designed to work effectively on each group.

Those criteria are qualitative and more or less subjective. To quantify those criteria, various validity indexes are commonly used.
\subsection{Validity Indexes}
Validity indexes (or indicators) are methods to quantify the quality of clustering results automatically. There are $2$ main types of validity indexes: internal and external. Internal indexes are calculated solely by the internal representation of the clustering results, while the external indexes compare the generated clusters with external grouping information. In this paper, we focus on the internal one since it is more useful in real application. After all, if we already have a good grouping information, we do not need to do the clustering anymore.

 In this paper, we experiment with four internal indexes, which are the most common for time-series clustering problems \cite{Chicco2003, Chicco2006, Chicco2012, Dent2012, Chicco2003tariff, Figueiredo2005, Tsekouras2007}. Due to the limited space, we only presented their general definitions here. For details and exact calculation, one could refer to \cite{davies1979cluster, Chicco2006}

\begin{itemize}
	\setlength\itemsep{0em}
	\item \textit{Clustering Dispersion Index} (CDI): this measures both the compactness of the clusters and how much each cluster differs from others. It is defined as the ratio of the mean of all clusters' intraset distances and the intraset distance of all centroids.
	\item \textit{Modified Dunn Index} (MDI): adapted from the original Dunn index. It is the ratio of maximum cluster intraset distance, and minimum distance between two cluster centroids.
	\item \textit{Davies-Bouldin Index} (DBI): measured by the average of the similarity between each cluster with its most similar cluster. The similarity between two clusters is defined by the ratio of the sum of intraset distances of two clusters to the distance between two centroids
	\item \textit{Mean Index Adequacy} (MIA): assesses the compactness of the clusters (i.e. the MIA is low if the cluster members are close together)
\end{itemize}

All these indexes have a common characteristic: the lower values they are, the higher quality they indicate. Note that these indexes are comparable only if the number of clusters are equal.

\section{Clustering Techniques}
\label{clustering}
In this paper, we investigated the RLP clustering ability of the following methods: \textit{K-Means} (KM), \textit{Self Organizing Map} (SOM), \textit{Hierarchical Clustering} (HC), and their variants.
\subsection{K-Means - KM}
The classical k-means (KM) algorithm \cite{tou1974pattern} follows a heuristic iterative procedure to group RLPs into $K$ clusters. It first picks $K$ cluster centroids (usually chosen randomly from RLPs set). Then each RLP is classified to its closest clusters (minimum distance between the RLP and the cluster centroid). The centroids are then recalculated by averaging their members' RLPs. The process is repeated until the cluster centroids are stable. In KM algorithm, the distance between two RLPs are measured by Euclidean distance. However, the studies in \cite{beyer1999nearest, Houle2010} suggest that cosine distance behaves better than Euclidean in high dimension space (i.e. more than $10$ features). Therefore, we also tested the spherical k-means (SKM) method where cosine dissimilarity is used.

\subsection{Self Organizing Map - SOM}
The Kohonen SOM \cite{SOM} is a special type of artificial neural network designed for unsupervised classification task. The SOM projects the input space into a reduced dimension space (usually a hexagonal bi-dimensional map), where proximity in input space is approximately preserved in output space. In general, the SOM may be considered a nonlinear generalization of \textit{Principal Component Analysis}~\cite{yin2008learning}, which is arguably the most popular linear dimensionality reduction tool. In this paper, we used a hexagonal bi-dimensional SOM as a dimensionality reduction tool. After a SOM model is trained on the RLPs, the obtained weight vectors of the SOM's units are grouped by k-means algorithm to obtain the final clusters. This combination of SOM and k-means is common in data mining on large dataset since it can overcome the \textit{dead units} problem when applying SOM alone~\cite{Figueiredo2005}.
\subsection{Hierarchical Clustering - HC}
There are two general types of hierarchical clustering (HC) \cite{anderberg2014cluster}: \textit{agglomerative}--`bottom up' and \textit{divisive}--`top down' approach. In this paper, the agglomerative one is employed, since it is more suitable for large datasets (divisive clustering with an exhaustive search is $O(2^n)$).

In agglomerative clustering, each RLP is initialized as a singleton cluster. Afterwards, these clusters are merged together in an iterative process, where in each step the two closest clusters are merged based on a specific \textit{linkage criterion}. The process continues until the desired number of clusters is achieved. The linkage criterion determines the similarity between two clusters. It is defined in term of pairwise distances between their members or centroids. The grouping result depends very much on the choice of the linkage criterion \cite{Chicco2012}. There are several available criteria such as \textit{complete}, \textit{single}, \textit{ward}, or \textit{average}.  We decided to experiment with ward, average, and single linkages, since they have unique characteristics and produce diverse clustering results, as shown in \cite{Chicco2003, Chicco2006, Tsekouras2007}. The three linkages and their characteristics are explained shortly as follows:
\begin{itemize}
	\setlength\itemsep{0em}
	\item \emph{Ward Linkage}: In ward linkage, the distance between two clusters are measured by the increase of within-cluster sums of squares (or other pairwise distance if the Euclidean is not being used) if the two clusters are merged. This criterion prefers small equal-size clusters over large ones.
	\item \emph{Average Linkage}: The average linkage criterion measures distance between two clusters by averaging pairwise distances between all pairs of the two clusters' members. This criterion tends to form large clusters of similar observations and put very dissimilar ones into small clusters.
	\item \emph{Single Linkage}: In single linkage, the distance between two clusters is equal to the distance between the closest pair of their members. This is an extreme criterion which usually leads to the formation of one large cluster together with many singleton or very small clusters.
\end{itemize}
A distinct advantage of HC is that any pairwise distance can be used independently with any linkage criterion. In this paper, we presented experiment results of HC using various combinations of Euclidean and cosine distance with different linkage criteria. We also tested HC using Minkowski distance of order $5$~\footnote{Euclidean distance is Minkowski distance of order $2$} with single linkage, since this combination produces superior result in \cite{Chicco2012}. The list of all experimented clustering methods with their abbreviations and detail descriptions are shown in Table \ref{tab:algorithms}.
\begin{table}[thb]
	\caption{Clustering methods with descriptions and properties explanations}
	\centering
	\small
	\begin{tabular}{p{1.5cm} p{7.5cm} p{6.5cm}}
		\hline Abbr. & Description & Properties \\ 
		\hline SOM & The Self-Organizing Map with hexagonal bi-dimensional map (10, 10) are used first for dimension reduction. The k-means clustering is then used to group weights of SOM nodes. & Produce a mix of big, medium, small and singleton clusters. Most of them has unique properties \\
		KM & The K-means clustering with Euclidean distance &  Tend to create equal-size medium clusters. Some clusters are not distinguishable\\
		SKM & The spherical K-means &  Similar to the KM method but tend to produce bigger clusters\\
		HC-W2 & The Hierarchical Clustering with Ward linkage criterion and Euclidean distance &  Tend to produce equal-size medium clusters with indistinguishable properties \\
		HC-S5 & The Hierarchical Clustering with Single linkage and Minkowski p = 5 & Produce one very big noisy cluster and many singleton or very small clusters\\
		HC-A2 & The Hierarchical Clustering with Average linkage criterion and Euclidean distance & Tend to create one big and quite noisy clusters, several small clusters with unique properties, and many very small or singleton clusters \\
		HC-SC & The Hierarchical Clustering with Single linkage criterion and Cosine distance & Similar to the HC-S5, but only produce singleton instead of very small clusters \\
		HC-AC & The Hierarchical Clustering with Average linkage criterion and Cosine distance & Similar to the HC-A2 but tend to produce bigger clusters\\
		\hline 
	\end{tabular} 
	\label{tab:algorithms}
\end{table}
\section{Experiment Results and Discussions}
\label{results}
\begin{figure}
	\centering
	\subfloat[HCS5\label{fig:HCS5}]{
		\includegraphics[width=.55\linewidth]{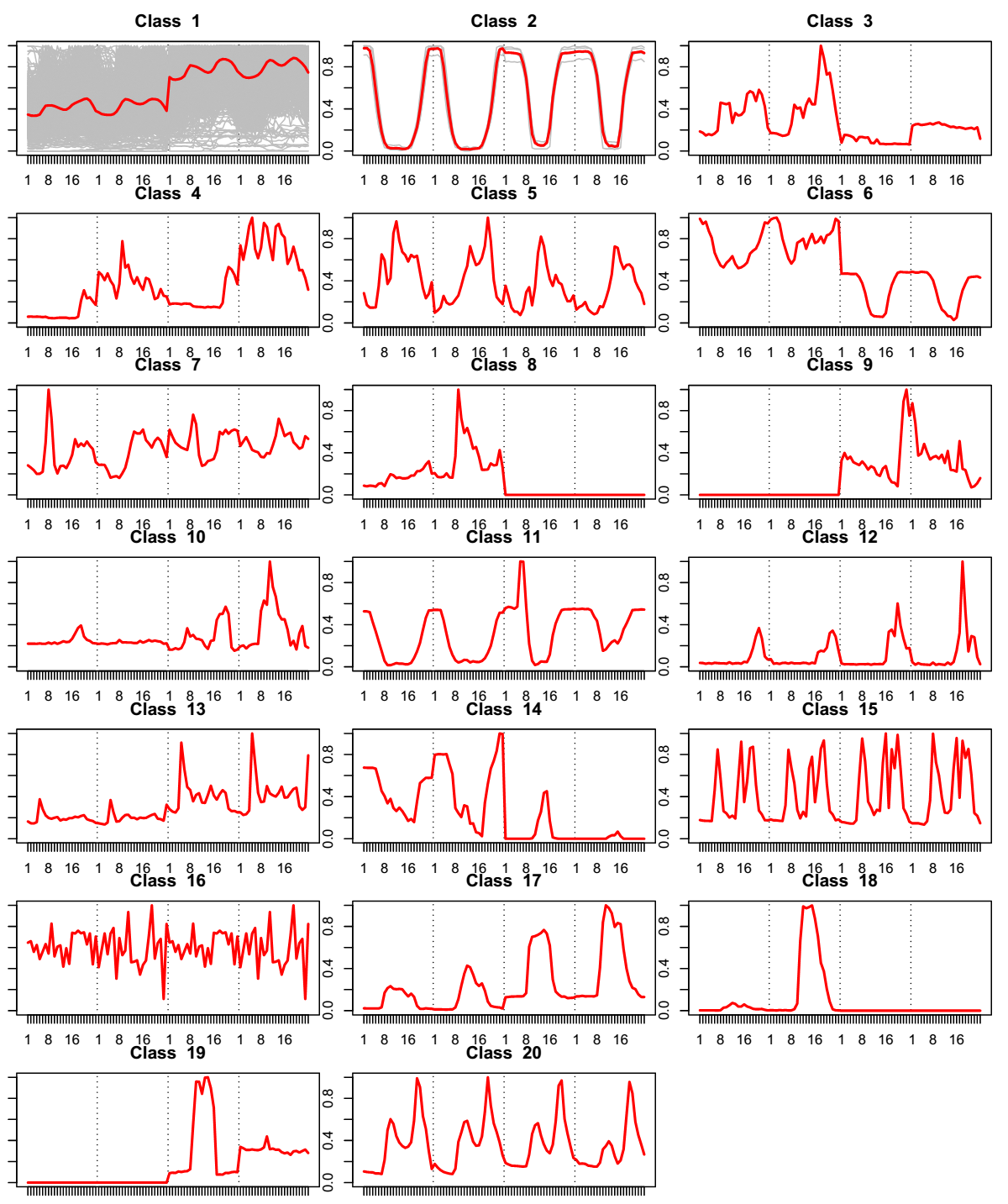}
	}
	\subfloat[HCAC\label{fig:HCAC}]{
		\includegraphics[width=.55\linewidth]{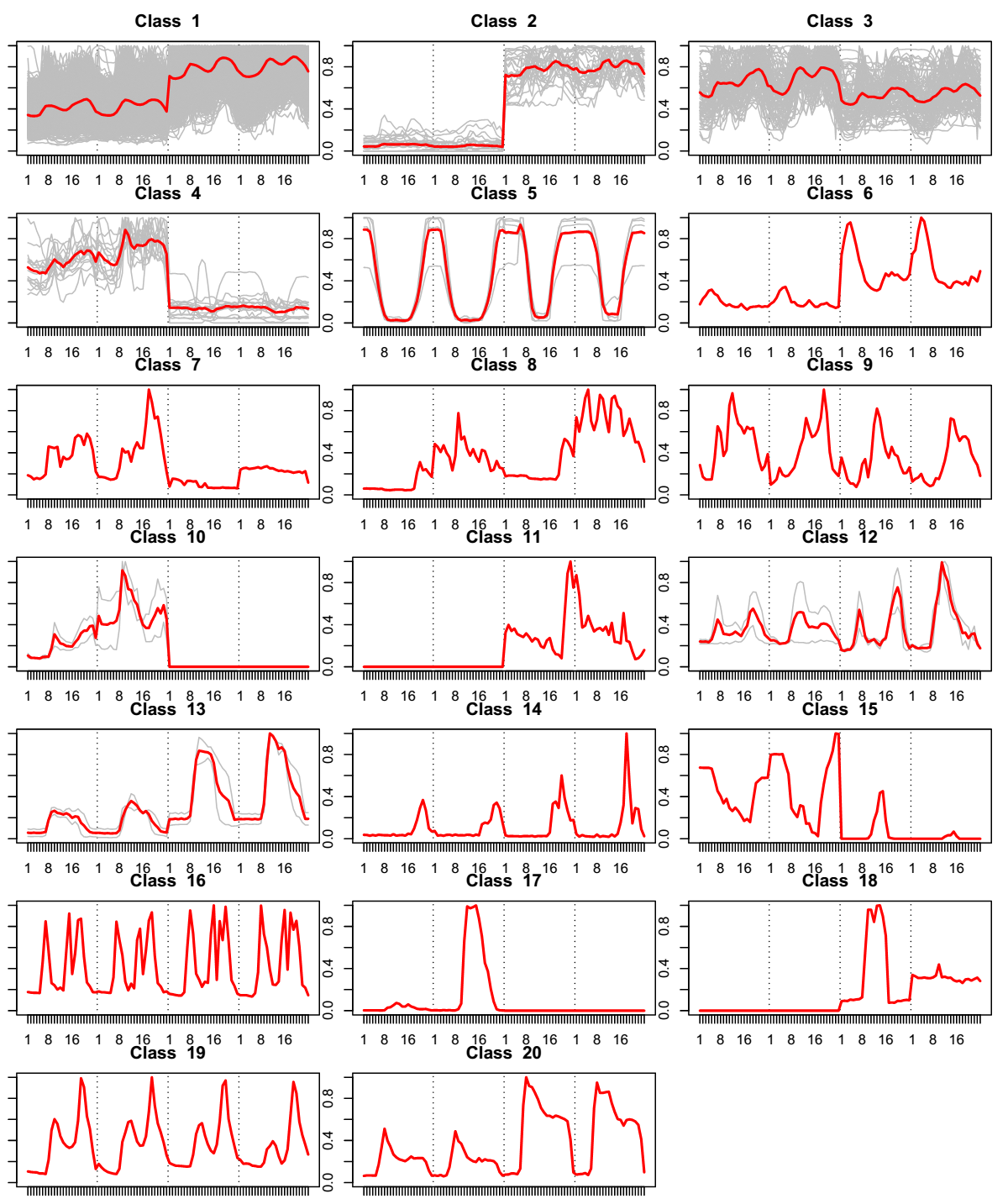}
	}

	\subfloat[SOM\label{fig:SOM}]{
		\includegraphics[width=.55\linewidth]{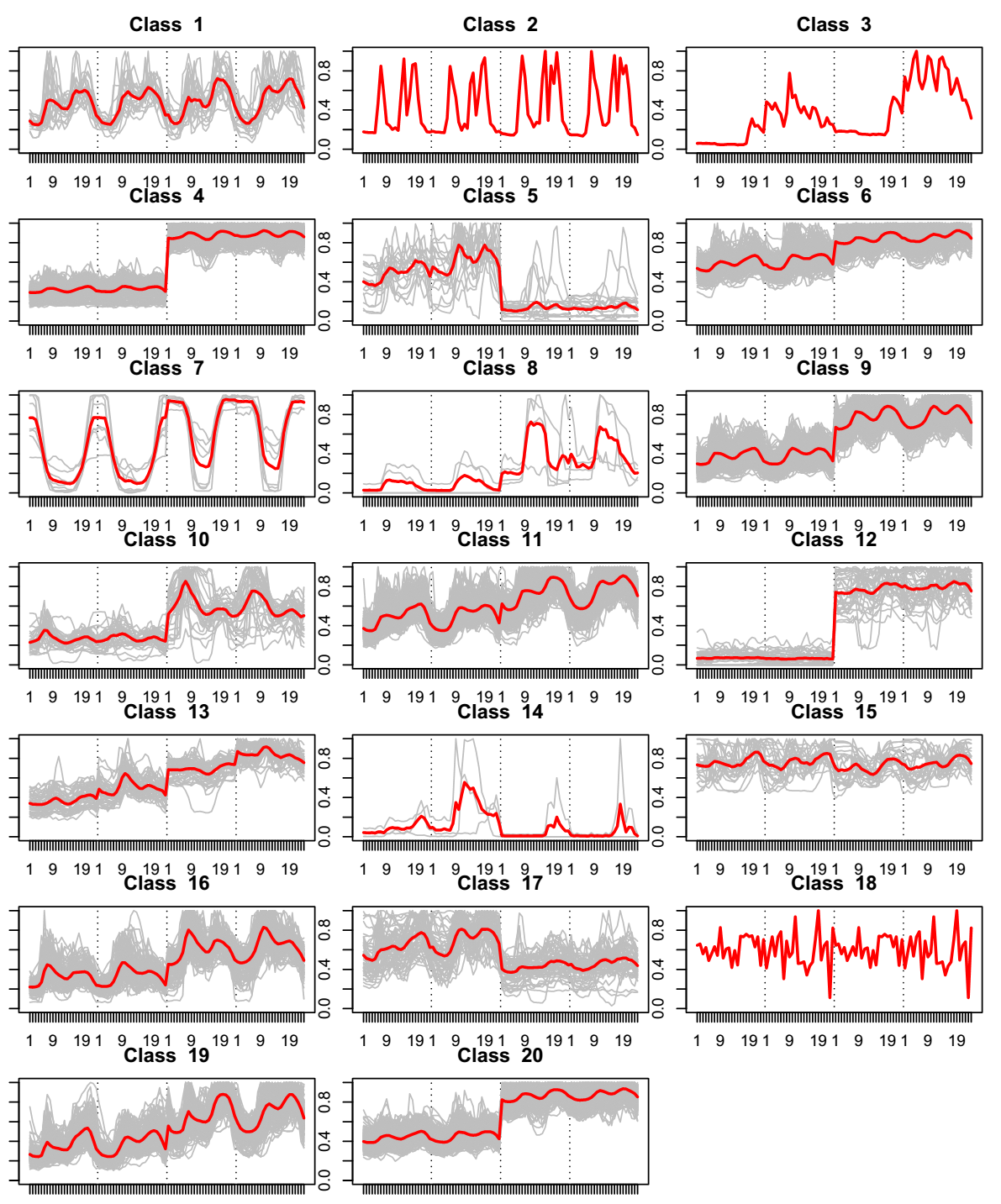}
	}
	\subfloat[SKM\label{fig:SKM}]{
		\includegraphics[width=.55\linewidth]{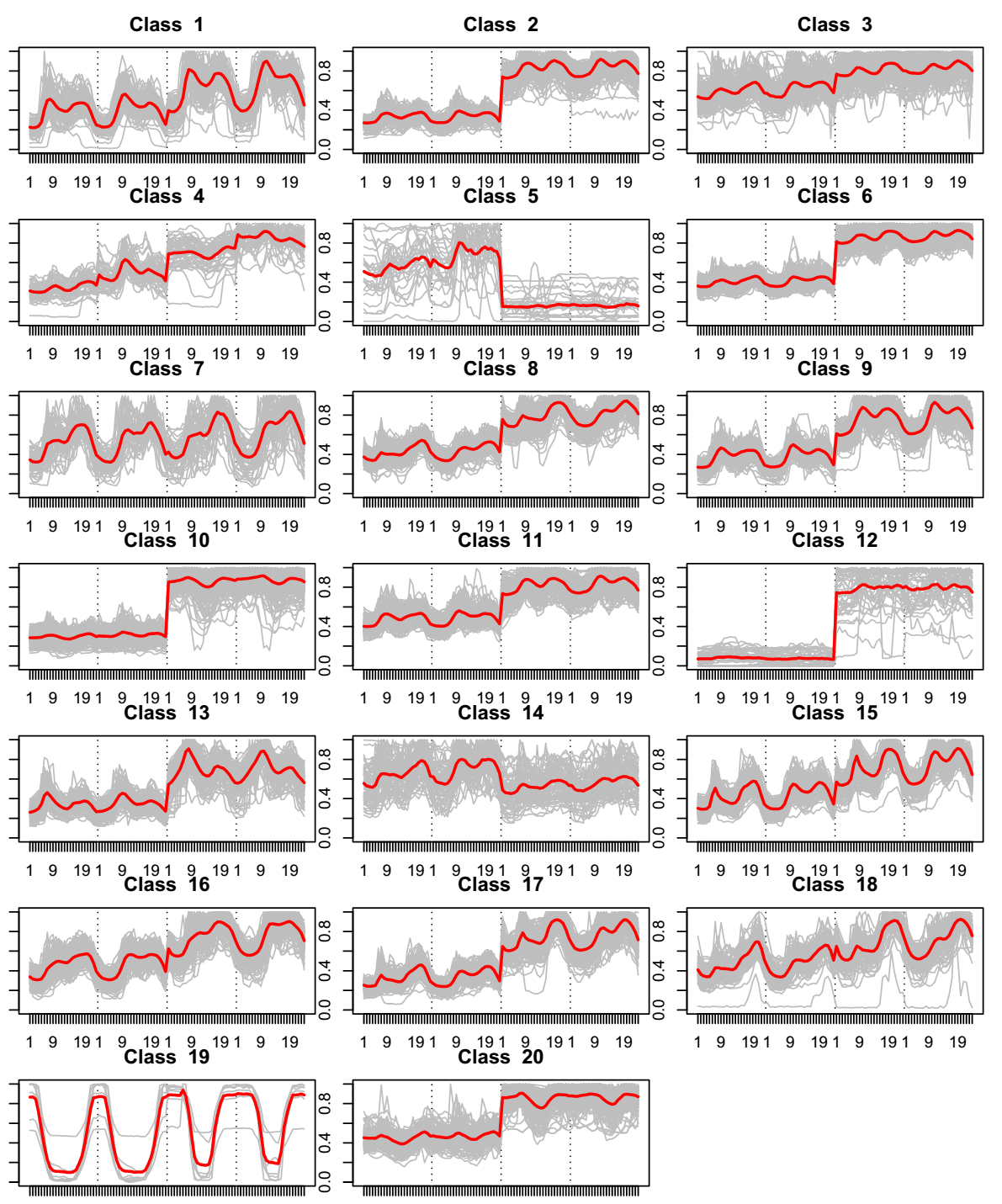}
	}
	\caption{Clustering Results of different methods when $K=20$. The red curves are the centroids of each cluster}
	\label{fig:clusteringresults}
\end{figure}

Figure \ref{fig:Validity} compares the validity indexes at different number of clusters using the $8$ methods listed in Table \ref{tab:algorithms}. The figure indicates that the HC methods with single linkage criterion (we use HC-Sx to indicate both HC-SC and HC-S5) gives seemingly the ``best'' clustering results. However, a completely different conclusion should be drawn when we look at the actual generated clusters presented in Figure \ref{fig:HCS5}. The HC-S5 method formulates one big cluster while putting outliers into singleton or very small clusters. It fails to cluster RLPs with low consumption during winter or summer (e.g. cabins), or RLPs which reacts to the season change differently. The only meaningful cluster is the second one, which seems to contain RLPs of lightings (off and on at the same time every day, and does not change by season). It is even worst in the case of HC-SC, where all the clusters are singleton except the biggest class (check the Table \ref{tab:memdist} to see member distribution of different clustering techniques). Despite of that poor results, all the validity indicators still strongly support HC-Sx. This observation casts doubt on the usefulness of these indicators, and poses questions on the previous studies that suggest using them to automatically select cluster results and/or determine the optimal number of clusters \cite{Tsekouras2007, Wijaya2014, Figueiredo2005, Chicco2003}.

The situation is better with HC methods using average linkage criterion (i.e. HC-Ax: HC-AC and HC-A2). Figure \ref{fig:HCAC} shows the clustering results of the HC-AC method, which has successfully identified four important consumers classes: 
\begin{itemize}
	\setlength\itemsep{0em}
	\item Cabins which are turned off during summer (class 2)
	\item Cabins which are turned off during winter (class 4)
	\item Lightings (class 5)
	\item Households that do not increase consumption in winter (class 3)
\end{itemize}
Most of the reasonable outliers recognized by HC-S5 are also successfully identified by HC-AC. Furthermore, HC-AC is able to correctly expand and merge singleton clusters of HC-S5. For example, class 10 of HC-AC contains all RLPs in class 8 of HC-S5, with one more similar RLP added. The same phenomenon happens to classes 12 and 13. Class 5 is the union of HCS5's class 11 and class 2. Despite its superior clustering result, all four validity indicators still prefer HC-Sx than HC-Ax. (HC-A2 produces similar but slightly worse result than HC-AC. Due to the limit in space, its result is not presented here).

The SOM method is able to identify even more interesting classes. Classes 2, 3 and 18 are the same outliers identified by the HC-S5 and HC-AC. However, they are all outliers with very unique and interesting properties. Class 2 has RLP with 3 daily peaks (instead of two); Class 3 are consumers who only consume during weekends; Class 18 looks like a white noise regardless of context, which we believe to be the smart home pilot with rooftop photovoltaic power station. Class 12 is basically class 2 of HC-AC, containing cabins which is off during summer, with $7$ more RLPs added to this class. Similarly, class 5 is class 4 of HC-AC with $4$ more RLPs, containing cabins which is off during winter; class 7 is class 5 of HC-AC with $3$ more RLPs, containing lighting consumptions. Some more classes with interesting properties are successfully identified by SOM method:

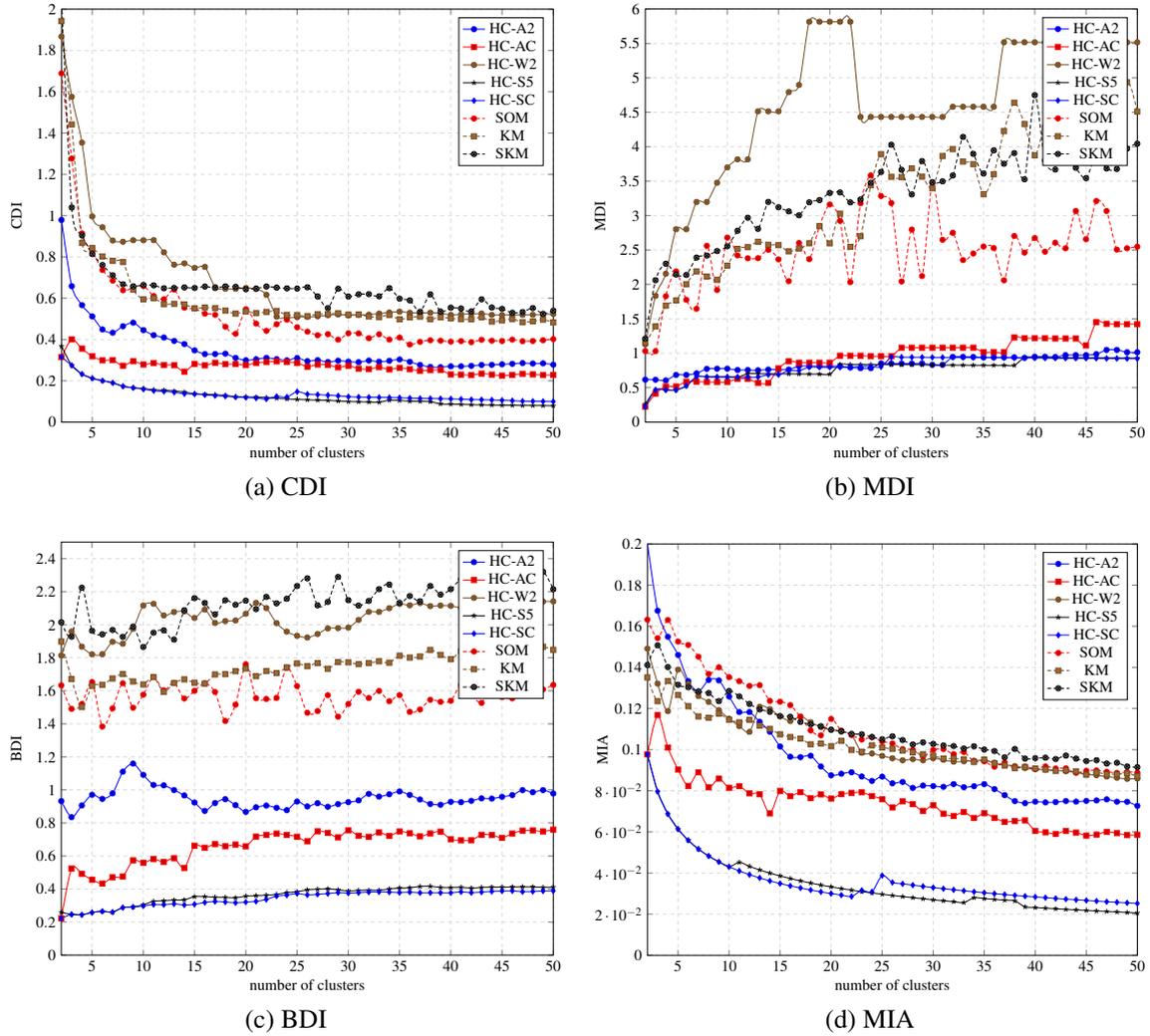
\begin{figure} [tb]
	\centering
	\subfloat[CDI\label{fig:CDI}]{
		\resizebox{0.5\linewidth} {!}{
			\begin{tikzpicture}
			\begin{axis}[
			width=\linewidth, 
			grid=major, 
			grid style={dashed,gray!30}, 
			xlabel=number of clusters, 
			ylabel = CDI,
			xmin=2, xmax=50,
			ymin=0, ymax=2, 
			smooth
			]
			\addplot
			table[x=nCluster,y=CDIHCA2,col sep=comma] {Results/ValidityResults.csv}; 
			\addlegendentry{HC-A2}
			\addplot
			table[x=nCluster,y=CDIHCAC,col sep=comma] {Results/ValidityResults.csv}; 
			\addlegendentry{HC-AC}
			\addplot 
			table[x=nCluster,y=CDIHCW2,col sep=comma] {Results/ValidityResults.csv}; 
			\addlegendentry{HC-W2}
			\addplot 
			table[x=nCluster,y=CDIHCS5,col sep=comma] {Results/ValidityResults.csv}; 
			\addlegendentry{HC-S5}
			\addplot 
			table[x=nCluster,y=CDIHCSC,col sep=comma] {Results/ValidityResults.csv}; 
			\addlegendentry{HC-SC}
			\addplot 
			table[x=nCluster,y=CDISOM,col sep=comma] {Results/ValidityResults.csv}; 
			\addlegendentry{SOM}
			\addplot 
			table[x=nCluster,y=CDIKM,col sep=comma] {Results/ValidityResults.csv}; 
			\addlegendentry{KM}
			\addplot 
			table[x=nCluster,y=CDISKM,col sep=comma] {Results/ValidityResults.csv}; 
			\addlegendentry{SKM}
			\end{axis}
			\end{tikzpicture}
		}
	}
	\subfloat[MDI\label{fig:MDI}]{
		\resizebox{0.5\linewidth} {!}{
			\begin{tikzpicture}
			\begin{axis}[
			width=\linewidth, 
			grid=major, 
			grid style={dashed,gray!30}, 
			xlabel=number of clusters, 
			ylabel = MDI,
			xmin=2, xmax=50,
			ymin=0, ymax=6,
			smooth
			]
			\addplot 
			table[x=nCluster,y=MDIHCA2,col sep=comma] {Results/ValidityResults.csv}; 
			\addlegendentry{HC-A2}
			\addplot
			table[x=nCluster,y=MDIHCAC,col sep=comma] {Results/ValidityResults.csv}; 
			\addlegendentry{HC-AC}
			\addplot 
			table[x=nCluster,y=MDIHCW2,col sep=comma] {Results/ValidityResults.csv}; 
			\addlegendentry{HC-W2}
			\addplot 
			table[x=nCluster,y=MDIHCS5,col sep=comma] {Results/ValidityResults.csv}; 
			\addlegendentry{HC-S5}
			\addplot 
			table[x=nCluster,y=MDIHCSC,col sep=comma] {Results/ValidityResults.csv}; 
			\addlegendentry{HC-SC}
			\addplot 
			table[x=nCluster,y=MDISOM,col sep=comma] {Results/ValidityResults.csv}; 
			\addlegendentry{SOM}
			\addplot 
			table[x=nCluster,y=MDIKM,col sep=comma] {Results/ValidityResults.csv}; 
			\addlegendentry{KM}
			\addplot 
			table[x=nCluster,y=MDISKM,col sep=comma] {Results/ValidityResults.csv}; 
			\addlegendentry{SKM}
			\end{axis}
			\end{tikzpicture}
		}
	}

	\subfloat[BDI\label{fig:BDI}]{
		\resizebox{0.5\linewidth} {!}{
			\begin{tikzpicture}
			\begin{axis}[
			width=\linewidth, 
			grid=major, 
			grid style={dashed,gray!30}, 
			xlabel=number of clusters, 
			ylabel = BDI,
			xmin=2, xmax=50,
			ymin=0, ymax=2.5,
			smooth
			]
			\addplot 
			table[x=nCluster,y=BDIHCA2,col sep=comma] {Results/ValidityResults.csv}; 
			\addlegendentry{HC-A2}
			\addplot
			table[x=nCluster,y=BDIHCAC,col sep=comma] {Results/ValidityResults.csv}; 
			\addlegendentry{HC-AC}
			\addplot 
			table[x=nCluster,y=BDIHCW2,col sep=comma] {Results/ValidityResults.csv}; 
			\addlegendentry{HC-W2}
			\addplot 
			table[x=nCluster,y=BDIHCS5,col sep=comma] {Results/ValidityResults.csv}; 
			\addlegendentry{HC-S5}
			\addplot 
			table[x=nCluster,y=BDIHCSC,col sep=comma] {Results/ValidityResults.csv}; 
			\addlegendentry{HC-SC}
			\addplot 
			table[x=nCluster,y=BDISOM,col sep=comma] {Results/ValidityResults.csv}; 
			\addlegendentry{SOM}
			\addplot 
			table[x=nCluster,y=BDIKM,col sep=comma] {Results/ValidityResults.csv}; 
			\addlegendentry{KM}
			\addplot 
			table[x=nCluster,y=BDISKM,col sep=comma] {Results/ValidityResults.csv}; 
			\addlegendentry{SKM}
			\end{axis}
			\end{tikzpicture}
		}
	}
	\subfloat[MIA\label{fig:MIA}]{
		\resizebox{0.5\linewidth} {!}{
			\begin{tikzpicture}
			\begin{axis}[
			width=\linewidth, 
			grid=major, 
			grid style={dashed, gray!30}, 
			xlabel=number of clusters, 
			ylabel = MIA,
			xmin=2, xmax=50,
			ymin=0, ymax=0.2,
			smooth,
			]
			\addplot
			table[x=nCluster,y=MIAHCA2,col sep=comma] {Results/ValidityResults.csv}; 
			\addlegendentry{HC-A2}
			\addplot
			table[x=nCluster,y=MIAHCAC,col sep=comma] {Results/ValidityResults.csv}; 
			\addlegendentry{HC-AC}
			\addplot 
			table[x=nCluster,y=MIAHCW2,col sep=comma] {Results/ValidityResults.csv}; 
			\addlegendentry{HC-W2}
			\addplot 
			table[x=nCluster,y=MIAHCS5,col sep=comma] {Results/ValidityResults.csv}; 
			\addlegendentry{HC-S5}
			\addplot 
			table[x=nCluster,y=MIAHCSC,col sep=comma] {Results/ValidityResults.csv}; 
			\addlegendentry{HC-SC}
			\addplot 
			table[x=nCluster,y=MIASOM,col sep=comma] {Results/ValidityResults.csv}; 
			\addlegendentry{SOM}
			\addplot 
			table[x=nCluster,y=MIAKM,col sep=comma] {Results/ValidityResults.csv}; 
			\addlegendentry{KM}
			\addplot 
			table[x=nCluster,y=MIASKM,col sep=comma] {Results/ValidityResults.csv}; 
			\addlegendentry{SKM}
			\end{axis}
			\end{tikzpicture}
		}
	}
	
	\caption{Validity Results (viewed best in color)}
	\label{fig:Validity}
\end{figure}

\begin{table}
	\centering
	\caption{Member distribution of different clustering techniques when $K = 20$}
	\resizebox{0.65\linewidth} {!}{
		\begin{filecontents*}{Results/MembersDistribution.csv}
Class,HC-SC,HC-S5,HC-AC,HC-A2,SOM,SKM,KM,HC-W2
1,3071,3068,2936,2894,21,84,117,236
2,1,4,26,40,1,237,104,283
3,1,1,90,14,1,184,106,78
4,1,1,15,47,322,50,399,195
5,1,1,5,18,18,26,362,203
6,1,1,1,49,205,400,6,171
7,1,1,1,4,8,56,223,267
8,1,1,1,5,4,269,226,191
9,1,1,1,1,862,218,38,154
10,1,1,2,4,32,235,241,186
11,1,1,1,2,392,277,170,229
12,1,1,2,1,33,40,309,148
13,1,1,2,3,48,106,259,66
14,1,1,1,2,3,70,103,189
15,1,1,1,1,25,199,119,146
16,1,1,1,1,124,177,163,18
17,1,1,1,1,54,195,29,148
18,1,1,1,1,1,100,47,100
19,1,1,1,1,225,7,49,66
20,1,1,1,1,711,160,20,16
\end{filecontents*}
\pgfplotstabletypeset[
		col sep=comma,
		string type,
		every head row/.style={before row=\hline,after row=\hline},
		every last row/.style={after row=\hline},
		]{Results/MembersDistribution.csv}
	}
	\label{tab:memdist}
\end{table}

\begin{itemize}
	\setlength\itemsep{0em}
	\item Classes that are not affected by season change (1, 15)
	\item Classes that greatly increase consumption in winters (4, 9, 10, 20)
	\item Classes that slightly increase consumption in winters (6, 11, 13, 19)
	\item Classes that decrease consumption in winters (17)
	\item Classes that do not vary much on daily usage (4, 6, 13, 15, 20)
	\item Classes that have breakfast peak greater than before-dinner peak (10, 16)
	\item Classes that have before-dinner peak greater than breakfast (1, 5, 11, 19)
\end{itemize}

The SKM method, as shown in Figure \ref{fig:SKM} generates several similar classes discovered by SOM. However, the clusters compactness and differentiability is worse. For example, class 2, 10, 12, and 20; or class 6, 8, 9, 11; they have indistinguishable properties. It fails to cluster RLPs with similar properties together. Instead, it tends to mix them into different clusters. The situation is similar in the case of HC-W2 and KM method (not shown here due to the space limit).

\section{Conclusions}
\label{conclusion}
In this paper, we proposed a new way to extract RLP features for consumers, which is suitable for the case study in Hvaler--Norway. We experimented with $8$ clustering methods. The results show that each method has different tendency on clustering outliers and forming classes. The HC-Sx methods tend to focus on isolating outliers into singleton or very small clusters, and group all other points into one single big cluster. On the other hand, the KM, SKM, and HC-W2 methods tend to mix outliers and ``fuzzy'' points into various indistinguishable classes. The SOM method produces the best result in general, as they isolate strong outliers into singleton classes, while grouping RLPs with similar unique properties together.

Four internal validity indicators, including CDI, MDI, BDI and MIA, were used to evaluate clusters' quality. The results suggest that these indicators tend to bias strongly towards isolating outliers, and do not penalize the formation of large and noisy group enough. Therefore, ones should be very careful when using these indicators to automatically select the optimal number of clusters or choose the best clustering method. We suggest the validity of clustering results should be judged manually and in context of the application or program that one wants to build. For example, ones usually need to consider the cost of mis-classification, the number of classes that could be served, or how big and distinct a class should be to be worth developing a dedicated program.
\bibliography{ConsumerClassification}
\bibliographystyle{abbrv}

\end{document}